# Soliton-induced liquid crystal enabled electrophoresis

Bing-Xiang Li,[1] Rui-Lin Xiao,[1] Sergij V. Shiyanovskii,[1] and Oleg D. Lavrentovich[1,2,*]

[1]Chemical Physics Interdisciplinary Program, Advanced Materials and Liquid Crystal Institute, Kent State University, Kent, Ohio, 44242, USA

[2]Department of Physics, Kent State University, Kent, OH, 44242, USA

* olavrent@kent.edu

**Abstract**

Manipulation of particles by a uniform electric field, known as electrophoresis, is used in a wide array of applications. Of especial interest is electrophoresis driven by an alternating current (AC) as it eliminates electrode blocking and produces a steady motion. The known mechanisms of AC electrophoresis require that either the particle or the surrounding medium are asymmetric. This asymmetry is usually assured before the field is applied, as in the case of Janus spheres. We report on a new mechanism of AC electrophoresis, in which the symmetry is broken only when the field exceeds some threshold. The new mechanism is rooted in the nature of electrophoretic medium, which is an orientationally ordered nematic liquid crystal. Below the threshold, the director field of molecular orientation around a spherical particle is of a quadrupolar symmetry. Above the threshold, the director forms a polar self-confined perturbation around the inclusion that oscillates with the frequency of the applied field and propels the sphere. The director perturbations are topologically trivial and represent particle-like solitary waves, called "director bullets" or "directrons". The direction of electrophoretic transport can be controlled by the frequency of the field. The AC directron-induced liquid crystal enabled electrophoresis can be used to transport microscopic cargo when other modes of electrophoresis such as induced charge electrophoresis are forbidden.

# I. INTRODUCTION

Electric field acting on a nematic liquid crystal produces a number of nonlinear nonequilibrium phenomena with a rich spectrum of spatiotemporal patterns in the director field $\hat{\mathbf{n}}(\mathbf{r},t)$ that specifies average local orientation of the molecules [1-3]. Among the most studied are one-dimensional (1D) and 2D director patterns [1-3]. Recently, 3D particle-like dissipative solitons, called "director bullets" that represent propagating solitary waves of self-trapped oscillating director driven by an alternating-current (AC) electric field, have been observed [4, 5]. The director bullets are topologically unprotected self-confined configurations that lack fore-aft [4] or left-right [5] symmetry. "Topologically unprotected" means that the self-confined configuration forms by a smooth director deformation from the uniform state; there is no topological charge associated with such a soliton. Since these formations are self-confined waves of the director field that survive collisions, an appropriate term for them is "directrons" that we use as a synonym of director bullets in what follows. Directrons propagate perpendicularly to the driving electric field $\mathbf{E}$ and leave the background director field $\hat{\mathbf{n}}_0 = \text{const}$ intact. The directrons exist in nematics with negative anisotropies of dielectric permittivity $\Delta\varepsilon = \varepsilon_\parallel - \varepsilon_\perp < 0$ and electric conductivity, $\Delta\sigma = \sigma_\parallel - \sigma_\perp < 0$ (the subscripts refer to the direction with respect to the director) [4, 5].

In this work, we demonstrate that the directrons can develop at colloidal spheres dispersed in a nematic with $\Delta\varepsilon < 0$ and $\Delta\sigma < 0$ that feature tangential orientation of the director $\hat{\mathbf{n}}$ at their surface. In absence of the electric field, the director field around these spheres is of a quadrupolar symmetry with two point defects-boojums at the poles [6], Fig. 1(a). Electrophoresis of these symmetric particles in an AC electric field is impossible. The so-called induced-charge electrophoresis that can transport metal-dielectric Janus spheres [7-10] is unable to cause a net displacement of a homogeneous sphere. The so-called liquid crystal-enabled electrophoresis (LCEP) that relies on the dipolar asymmetry of the director configuration around particles that exists prior to the electric field application [11-13] is also ineffective because $\hat{\mathbf{n}}(\mathbf{r},t)$ around a tangential sphere is of a higher quadrupolar symmetry. Electrically induced directron dresses around the spheres, however, bring about a necessary polar symmetry in the structure and render the tangentially anchored spheres electrophoretically active under the AC field. The structure of the directrons that form above some electric field threshold is similar to the directrons described for uniformly aligned nematics without colloids [4, 5]. The directron-dressed spheres move in the plane perpendicular to $\mathbf{E}$; depending on the frequency or amplitude of the field, the spheres can move parallel, perpendicularly or at some angle to the uniform background director $\hat{\mathbf{n}}_0$. The soliton-dressed particles survive head-to-head collisions with each other, restoring their mobility. The effect can be used for electrically controllable transport of microcargo when other mechanisms of electrophoresis, such as linear electrophoresis, induced charge or liquid crystal enabled electrophoresis are ineffective. Since the motility of the spheres requires a formation of the directrons, we call the effect a directron-induced liquid crystal enabled electrophoresis (DI-LCEP).

## II. MATERIALS AND EXPERIMENTAL DESIGN

We used a nematic liquid crystal 4'-butyl-4-heptyl-bicyclohexyl-4-carbonitrile (CCN-47, purchased from Nematel GmbH). The material is of the $(-,-)$ type, with a negative anisotropy of both permittivity and conductivity. To confirm this classification, we measured the conductivity and permittivity of CCN-47 by using an LCR meter 4284A (Hewlett-Packard) and cells with planar (alignment agent polyimide PI-2555, HD MicroSystems) and homeotropic (polyimide SE1211, Nissan) alignment at $55°C$, $\sigma_\parallel \approx 0.9 \times 10^{-8} \, \Omega^{-1} \text{m}^{-1}$, $\sigma_\perp \approx 1.0 \times 10^{-8} \, \Omega^{-1} \text{m}^{-1}$, $\varepsilon_\parallel = 4.9$ and $\varepsilon_\perp = 8.2$. The cell is composed of two glass substrates coated with indium tin oxide (ITO), which serve as the transparent electrodes of active area $5 \times 5 \, \text{mm}^2$. The alignment layers PI-2555 coated on the surface of ITO, followed by 5 minutes soft baking at $90°C$ and then one hour hard baking at $275°C$. The PI-2555-coated substrates are rubbed in an antiparallel fashion to provide a planar orientation of the director in the $xy$ plane of the cell. The cell thickness is $8 \, \mu\text{m}$ or $20 \, \mu\text{m}$. As tangentially anchored colloids, we used polystyrene spheres (density $1.06 \, \text{g/cm}^3$) of diameter $2R = 1.5 - 8.5 \, \mu\text{m}$. The dispersion of spheres in CCN-47 is ultrasonicated for one hour before filling the cell at elevated temperatures at which CCN-47 is in the isotropic phase. The concentration of polystyrene spheres was small, 0.1% or less by volume, in order to avoid collective effects. Note, however, that since the colloids are moved by the directrons, we could still observe pairwise collisions, as described later.

The initial director $\hat{\mathbf{n}}_0 = (0,1,0)$ is along the $y$ axis. The applied AC electric field of frequency $f = 1-1000$ Hz is normal to the substrate, $\mathbf{E} = (0,0,E)$. The temperature of the cells was controlled with a Linkam LTS350 hot stage and a Linkam TMS94 controller. The AC voltage was applied using a waveform generator (Stanford Research Systems, Model DS345) and an amplifier (Krohn-hite Corporation, Model 7602).

The experimental images were taken by using polarizing Nikon TE2000 inverted microscope equipped with two cameras: Emergent HR20000 with frame rate up to 1000 fps and MotionBLITZ EOSens mini1 (Mikrotron GmbH) with frame rate up to 8000 fps. The location of colloidal particles was tracked by an open-source software ImageJ and its plugin TrackMate [14]. The velocities of particles were obtained by measuring the $x, y$ coordinates of the particles as a function of time.

## III. EXPERIMENTAL RESULTS
### A. Electric field induced cargo transportation

In absence of the field, director deformations around a tangentially anchored sphere are quadrupolar, extending over distances comparable to the radius of the sphere $R$, Fig. 1(a) [6]. Once the AC field of a fixed frequency $f$ exceeds some threshold $E_{th}$, the sphere acquires an asymmetric director "dress" of dipolar symmetry that extends over much larger distances $\sim (4-8)R$. The dressed spheres start to move in the plane of the cell, Figs. 1-4. The particles move without stopping until they reach the edge of electrodes. The translation distance can be 5 mm or larger (depending on the electrode size) which is 1000 times the diameter of the particle. The director structures that appears above $E_{th}$ and the direction of motion depend on the frequency and amplitude of the applied field. In particular, the director fields around the spheres at high frequences $f = (90-700)$ Hz are similar to the so-called directrons $\mathrm{B}_{90}^{h}$ described in Ref. [4], while the dresses forming at $f = (5-40)$ Hz are similar to the directrons of $B_\alpha^l$ described in Ref. [5]. Below we describe the details of the high-frequency, Figs. 1, 2, and the low-frequency, Figs. 3,4, regimes.

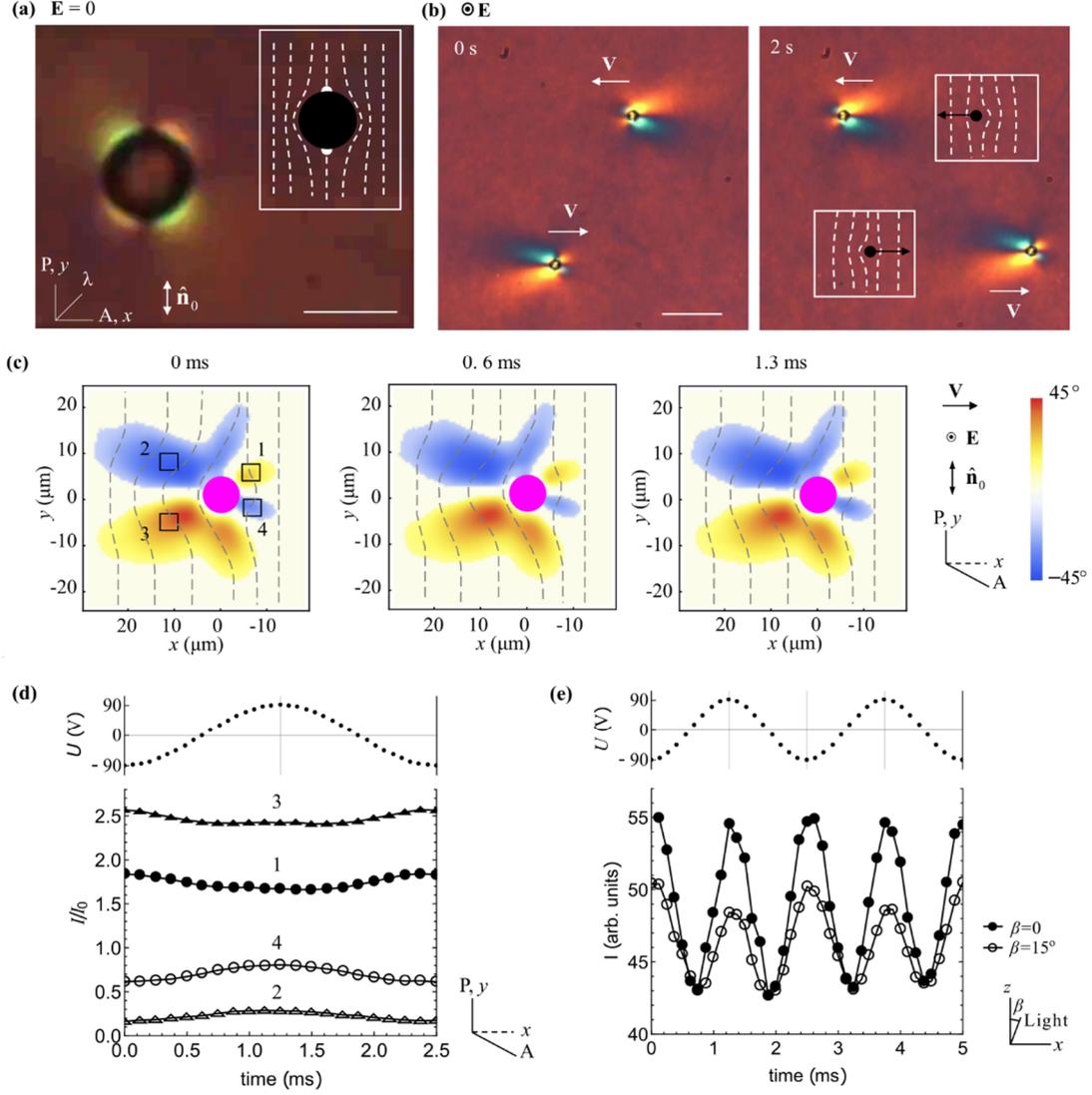

**FIG. 1.** Director field around tangentially anchored polystyrene spheres of diameter $2R = 8\,\mu m$ in CCN-47 cells of thickness $d = 20\,\mu m$, at $T = 55°C$. (a) Polarizing microscope image of the quadrupolar director field around the sphere in the field-free state. Scale bar $10\,\mu m$. (b) Spheres acquire $B_{90}^h$ directron dresses and mobility under an AC voltage $U = 42$ V of frequency $f = 150$ Hz. Scale bar $50\,\mu m$. (c) Maps of the in-plane director distortions around moving spheres ($U = 63$ V, $f = 400$ Hz) deduced from polarizing microscopy with two polarizers crossed at $60°$. The time step between the images is $1/4$ of the voltage period, with "0 ms" corresponding to the negative

extremum of the voltage. (d) Time/voltage dependence of the transmitted light intensity averaged over the areas 1, 2, 3, and 4 shown in (c). (e) Time/voltage dependence of transmitted light intensity at the location 3 for crossed polarizers at normal, $\beta = 0$, and oblique incidence, $\beta = 15°$.

## B. Director structure of directrons in high-frequency field

The director field within the field-induced dress lacks the left-right symmetry with respect to $\hat{\mathbf{n}}_0$, Fig. 1(c). To decipher the director details, we used a full wave (530 nm) optical compensator with the optic axis aligned under $45°$ to the polarizer ($y$-axis) and analyzer ($x$-axis), Fig. 1(a). In this setting, regions with a uniform background director $\hat{\mathbf{n}}_0 = (0,1,0)$ appear red. The regions in which the actual director deviates from the $y$-axis in an anti-clockwise manner appear yellow, while the areas with a clockwise director tilt are blue, Fig. 1(a-c).

The director shows a dynamic behavior, oscillating with the same frequency as the frequency of the applied AC electric field, Fig. 1(c-e). The dynamics of in-plane deformations was established by observations between the polarizer and analyzer decrossed at $60°$, according to the protocol described in Refs. [4, 5]. The in-plane azimuthal distortions do not change their curvature when the polarity of the voltage is reversed. To determine the period and polarity of out-of-plane director oscillation, we used oblique propagation of light [4, 5]. The cell is tilted so that the light beam of the polarizing optical microscope enters the cell at the angle $\beta = 15°$ from the normal to the cell [4, 5]. The dynamics of light intensity suggest that the polar tilt $\theta$ oscillates in phase with the applied voltage. The overall director configuration and dynamics in the high-frequency dresses are thus similar to that of the director in $B_{90}^h$ directrons reported in Ref. [4]. Because of that, we call a tangential sphere with a directron dress induced by the high frequency field a $B_{90}^h$-dressed colloidal sphere. Despite the noted similarities in the director fields, many properties of the spheres dressed in $B_{90}^h$ directrons and the particle-free pure $B_{90}^h$ directrons are different.

First, the threshold field at which the directrons appear around the spheres is significantly lower than the threshold field of $B_{90}^h$ directrons in a cell of the same thickness addressed by the field of the same frequency $f$. For example, for $d = 8\,\mu m$ and $f = 100\,Hz$, $E_{th} = 0.55\,V/\mu m$ for spheres of a diameter $2R \approx 1.5\,\mu m$, Fig. 2, while the colloid-free cells show a much higher threshold, $E_{th} \approx 5\,V/\mu m$ for directron appearance. Furthermore, the colloid-free samples exhibit a very narrow field range of $B_{90}^h$ existence, typically within $(1.0-1.1)E_{th}$ for a given $d = const$, $f = const$. In cells with colloids, the range of stability of $B_{90}^h$-dressings is substantially expanded, to $(1.0-2.5)E_{th}$. Above $2.5E_{th}$, the field causes an electrohydrodynamic instability in the entire area of the cell with colloids. Within the range $(1.0-2.5)E_{th}$, the speed of $B_{90}^h$-dressed spheres grows with the square of the field, $v = a(E^2 - E_{th}^2)$, where $a = (10.4 \pm 0.4) \times 10^{-18}\,m^3 V^{-2} s^{-1}$ is a nonlinear mobility, Fig. 2. The speed is on the order of $10\,\mu m/s$ which is smaller than typical velocity of sphere-free $B_{90}^h$ directrons, $(400-1200)\,\mu m/s$ in a similar $8\,\mu m$ cell. In thicker cells, $d = 20\,\mu m$, the speed of $B_{90}^h$-dressed spheres of diameter $2R \approx 8\,\mu m$ driven by the field $E \approx 6\,V/\mu m$, $f = 700\,Hz$, is $33\,\mu m/s$, which is again noticeably smaller than the speed $300\,\mu m/s$ of particle-free directrons in a similar cell. The exact dependencies of the speed on parameters such as $R, d, U, f$ and temperature require further studies, which are complicated by the finite range of the driving parameters $U$ and $f$, by fine balance of the surface anchoring and elasticity of the director around the spheres that depends on $R$ and temperature, and by the possible changes of ionic content of the system upon addition of colloids.

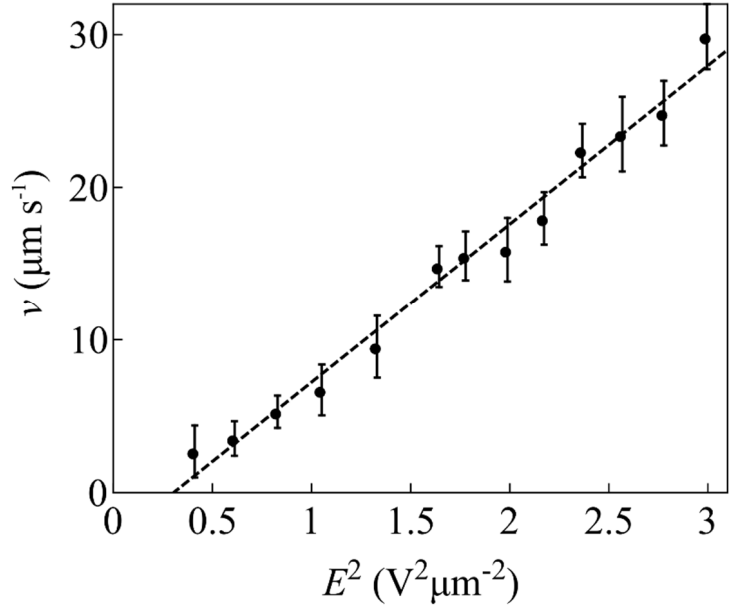

**FIG. 2.** Voltage dependence of the speed of $B_{90}^{h}$-dressed polystyrene sphere in CCN-47 at $T = 55°C$ ; $2R = 1.5\,\mu m$ , $d = 8\,\mu m$, $f = 100\,Hz$ . Data averaged over 10 spheres. The error bars show the standard deviation. The dashed line is a linear fitting which yields $E_{th} \approx 0.55\,V/\mu m$.

### C. Director structure of directrons in low-frequency field

At low frequency, the spheres acquire a directron dress similar to that of the previously described $B_{\alpha}^{l}$ directrons [5]; the subscript is the angle between the background director $\hat{\mathbf{n}}_0$ and the velocity vector. The principal difference between the $B_{90}^{h}$ and $B_{\alpha}^{l}$ directrons is that the $B_{90}^{h}$ directrons are comprised of two main sectors of the director deformations (director tilts to the right in one segment and to the left in the other segment) [4], while in the $B_{\alpha}^{l}$ directrons, there are four segments of the director tilt of comparable amplitude [5]. The analysis of the director field in $B_{\alpha}^{l}$ dressings is presented in Figs. 3 and 4. These two figures illustrate two cases of $B_{\alpha}^{l}$ dressings, namely, $B_{90}^{l}$, Fig. 3 and $B_{15}^{l}$, Fig. 4.

In the $B_{90}^{l}$ dresses, the in-plane director tilts in segments 1, 2, 3, and 4 oscillate, changing their polarity with the frequency $f$, as evidenced by observations with decrossed polarizers, Fig. 3(b,c). The field-induced director deformations preserve mirror symmetry with respect to a plane perpendicular to $\hat{\mathbf{n}}_0$, but lack it along $\hat{\mathbf{n}}_0$, Fig.

3(b,c). As a result, a sphere dressed in a $B_{90}^l$ directron moves perpendicularly to $\hat{\mathbf{n}}_0$, Fig. 3(a). The dynamics of light intensity measured from the normal $\beta = 0$ and oblique $\beta = 15°$ incidences according to the protocol described in Ref. [5], bring into evidence that the polar director tilt oscillates in phase with the applied voltage, Fig. 3(d). This director dynamics is thus similar to that of the director inside the particle-free $B_{90}^l$ directrons at low frequency reported in [5].

The tilt angle $\alpha$ of trajectories of the $B_\alpha^l$-dressed spheres can be changed by the driving voltage. As $U$ decreases below 13 V, at a fixed $f = 20\,\text{Hz}$, one observes spheres moving obliquely to $\hat{\mathbf{n}}_0$, $\alpha < 90°$. An example with $\alpha = 15°$ and $B_{15}^l$ dress is shown in Fig. 4. In this structure, the director oscillations are similar to those in Fig. 3, i.e., the azimuthal tilts change their polarity with the frequency $f$. The principal difference is that the structure shows no mirror symmetry with respect to any plane perpendicular to the cell, Fig. 4(b). The reason for the asymmetric structure is not clear, but can be tentatively associated with the increased role of surface anchoring and its plausible inhomogeneities at the surface of the spheres once the field becomes weaker; shape deviations from an ideal sphere might also be of importance.

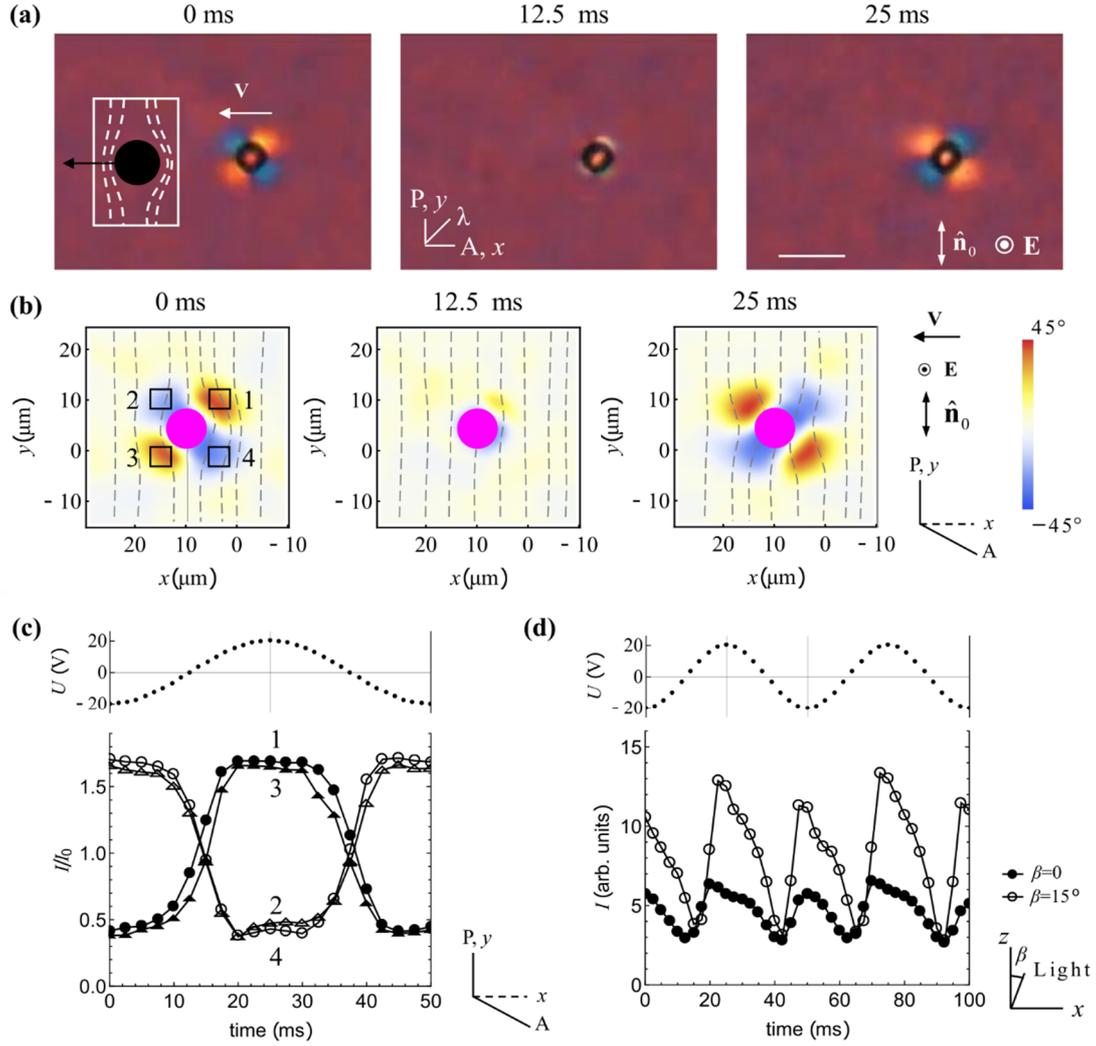

**FIG. 3.** Polystyrene spheres dressed in $B_{90}^l$ directron in CCN-47 ; $2R=8\,\mu m$, $d=20\,\mu m$, $T=55°C$, $f=20\,Hz$, $U=14\,V$. (a) Polarizing microscopy with wave plate demonstrating oscillating in-plane textures and mobility perpendicular to the background director. Scale bar $20\,\mu m$. (b) Maps of the in-plane director distortions reproduced from polarizing microscopy with two polarizers crossed at $60°$. The time step between images is a quarter of the voltage period, with "0 ms" corresponding to the negative extremum of the voltage. (c) Time/voltage dependence of light intensity averaged over the areas 1, 2, 3, and 4 shown in (b). (d) Time/voltage dependence of transmitted light intensity at the location 3 for the crossed polarizers at normal, $\beta=0$, and oblique, $\beta=15°$, incidence.

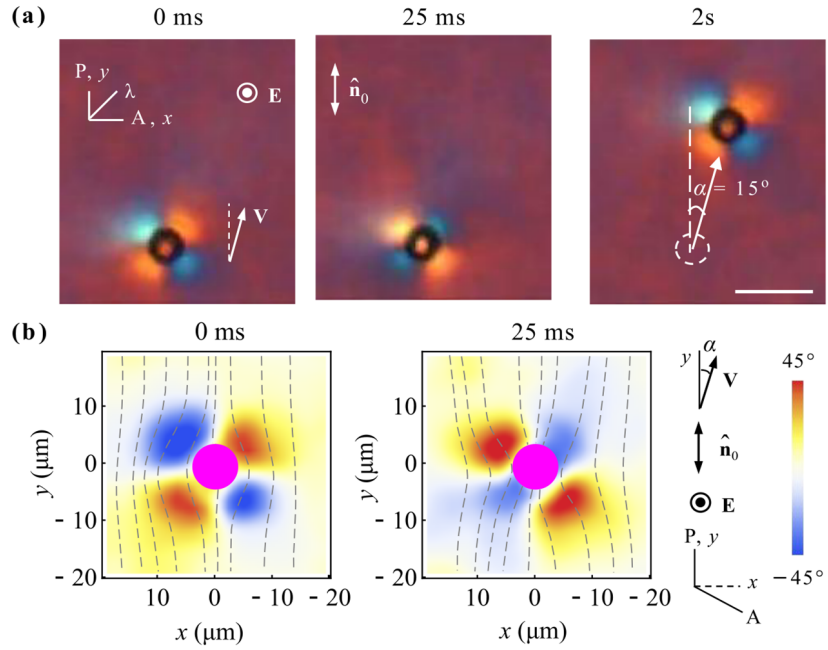

**FIG. 4.** Polystyrene sphere dressed in $B^l_{15}$ directron in CCN-47, $2R = 8\,\mu m$, $d = 20\,\mu m$, $T = 55°C$, $f = 20\,Hz$, $U = 12.5\,V$. (a) A sphere acquires a directron dress and mobility. Scale bar $20\,\mu m$. (b) In-plane director distortions around the sphere. The time step between images is half of the voltage period, with "0 ms" corresponding to the negative extremum of the voltage.

Figure 5 shows the voltage-controlled "phase diagrams" of colloid-free directrons and directrons formed around colloids, for driving frequencies 20 Hz, Fig. 5(a), and 500 Hz, Fig. 5(b). In general, the directrons dressing colloids exist in a wider voltage range, especially in the case of high frequency driving, Fig.5 (b).

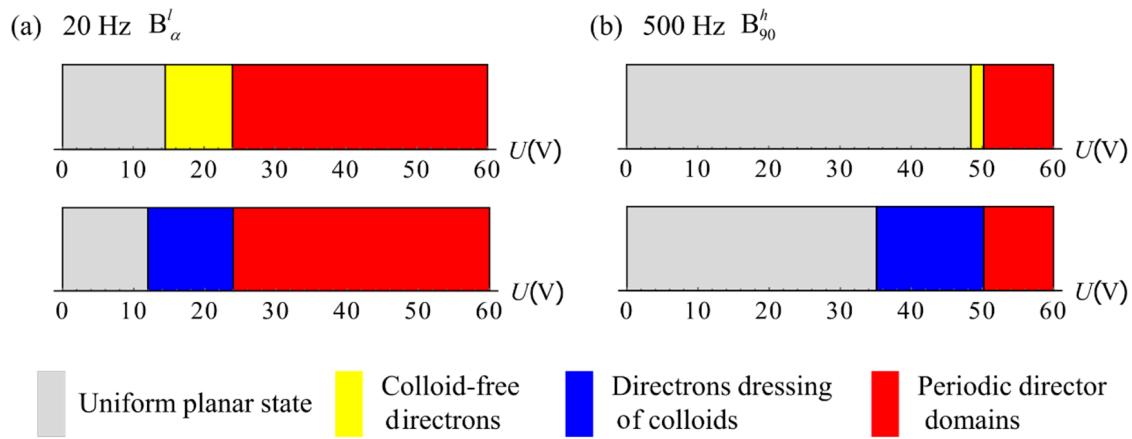

(a) 20 Hz $B_\alpha^l$

(b) 500 Hz $B_{90}^h$

Uniform planar state | Colloid-free directrons | Directrons dressing of colloids | Periodic director domains

**FIG. 5**. Phase diagrams of colloid-free directrons and directrons formed around colloids in CCN-47 at the low frequency 20 Hz (a) and the high frequency 500 Hz (b), $2R = 8\,\mu m$, $d = 20\,\mu m$, $T = 55°C$.

### D. Collisions of two directron-dressed spheres

It is commonly known that solitons can survive collisions and restore their shape and propagation mode in head-to-head encounters [15]. The same is true for the standing-alone directrons [4, 5]. This feature is the ultimate reason for the term "soliton", as it stresses particle-like properties of the solitary waves [16]. The directron-dressed spheres in our experiments show a similar ability to survive collisions and restore their dresses, even when they collide head-to-head. Since the solid particles cannot penetrate each other, the scenarios of encounters are very peculiar, as illustrated in Fig. 6 in which two $B_{90}^h$-dressed spheres move towards each other. Their initial impact distance $\Delta y$ (the separation of the centers of mass along the $y$-axis) is small, $0.5R$, Fig. 6(a, c). As the spheres approach each other, they slow down and $\Delta y$ decreases to zero, but once their centers arrive at the same $x$-coordinate, the $\Delta y$ distance increases to about $4R$. The effective repulsion along the $y$-axis is caused by impermeability of the particles and by elastic repulsion between their soliton dresses. Remarkably, after the spheres part with each other, they completely restore the soliton dresses, speed and horizontal direction of propagation, Fig. 6.

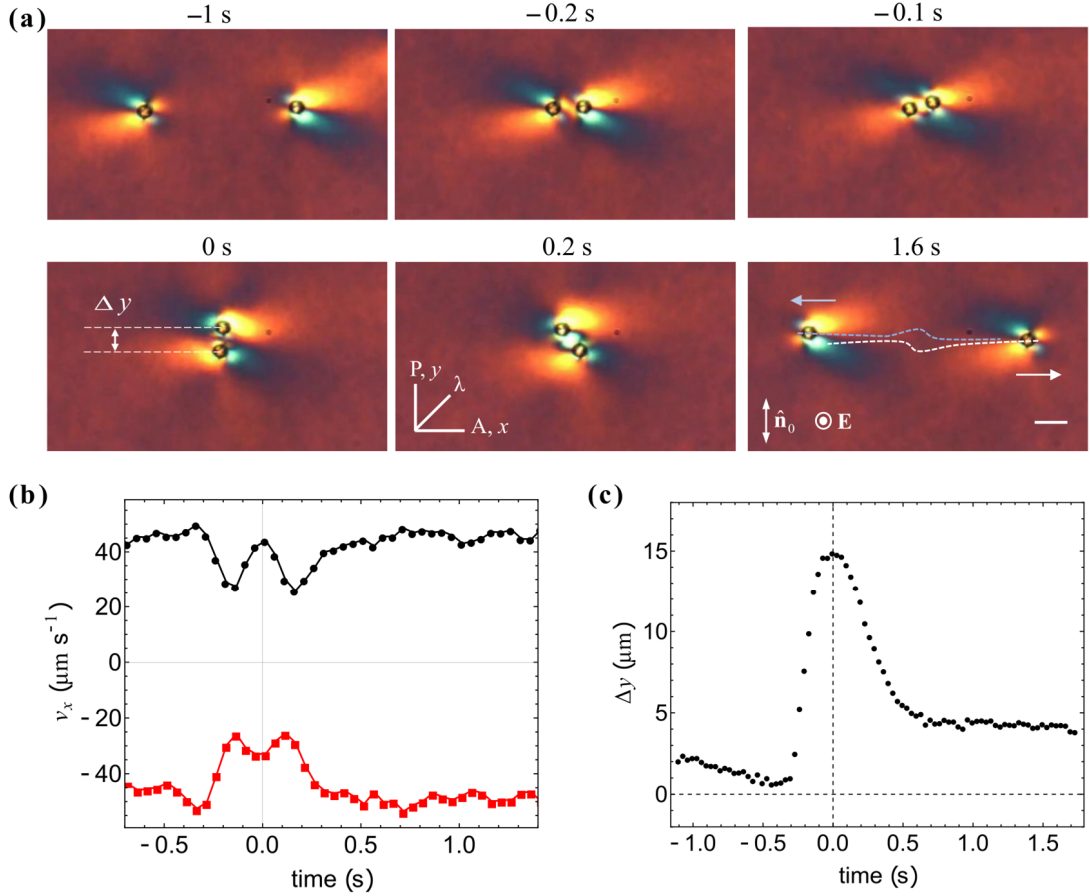

**FIG. 6.** Collision of two particles dressed in the $B_{90}^h$ directrons in CCN-47, $2R = 8\,\mu m$, $d = 20\,\mu m$, $T = 55°C$, $f = 150\,Hz$, $U = 42\,V$. (a) Polarizing microscope textures of the approach, collision and recovery. Scale bar $20\,\mu m$. (b) Time dependence of the velocity $v_x$. (c) Separation of particles along the $y$-axis as a function of time.

## IV.   DISCUSSION

It is known that colloidal particles placed in a liquid crystal electrolyte can become mobile when the director field around them is of a dipolar symmetry. The effect is called liquid crystal-enabled electrophoresis or LCEP [11-13, 17-18]. LCEP of spheres in nematic is effective when the director is anchored perpendicularly to the surface of a sphere and the cell thickness is significantly larger than the diameter of the particle. In this case, the director field acquires a dipolar symmetry, representing a locally radial structure in the vicinity of the sphere and a topological defect, the so-called hyperbolic hedgehog, next to it. In presence of the electric field, this dipolar structure separates electric charges that cause directional propulsion of the colloid with the velocity growing as $E^2$, so that the effect can be driven by an AC field [11, 12, 17, 18]. Nonlinear electrophoresis can also be caused by a pulsed high-frequency AC electric field that couples dielectrically to the dipolar director around a perpendicularly anchored sphere [19]. In the case of a tangentially anchored sphere, however, these mechanisms are not valid, as the director and charge separation patterns are of a quadrupolar symmetry with two planes of mirror symmetry, one parallel to the bounding plates and one normal to $\hat{\mathbf{n}}_0$. The present work shows that the tangentially anchored spheres become electrophoretically active through formation of electrically-triggered directron dresses around them. These directron dresses are similar to the 3D particle-like solitary waves called directrons and described earlier for high [4] and low [5] frequencies of an electric field acting on a uniformly aligned nematic. The speed of spheres grows with the square of the field, similarly to the conventional LCEP, but with that difference that the LCEP shows no threshold behavior while the effect described in this work does show a threshold behavior. Given all these similarities and differences, we call the observed phenomenon a directron-induced LCEP, or DI-LCEP.

In the description above, we presented the data for two different geometries, exploring collidal spheres of diameter 1.5 μm in 8 μm cell and of diameter 8 μm in 20 μm cell. The smaller particles allows one to obtain a better statistics on the propulsion speed, Fig. 2. The larger particles, on the other hand, are better suited to establish the director structure around the spheres, Figs. 1,3-6. We determined experimentally that the ratio of the sphere diameter to the cell thickness $2R/d$ should be in the range from 0.15 to 0.4 for the directron to form around the spheres. The effect of $2R/d$ on the stability of sphere-triggered directrons can be qualitatively explained as follows. As demonstrated in the previous work on colloid-free directrons, [4,5], the length $l$ and the width $w$ of directrons are related to the cell thickness $d$, namely, $l \approx (2-5)d$, $w \approx 2d$. Since the directron around a colloidal sphere is of a similar size as the colloid-free directrons, the ratio $2R/d$ is expected to be in a specific range, as observed. If the colloidal sphere is too big, it over-stretches the director deformations beyond the length-scale that corresponds to a stable self-confinement. If the sphere is too small, there are two reasons why the directrons do not dress them. First, a small sphere does not modify substantially the director field of the directron which is of a typical size $2d$. Second, if the particle is smaller than the de Gennes-Kleman anchoring extrapolation length $K/W$, where $K \sim 10$ pN is the typical elastic constant and $W \sim 10^{-4} - 10^{-6}$ J/m$^2$ is the polar anchoring coefficient [20], its surface anchoring is not strong enough to produce substantial director deformations.

As compared to the induced-charge electrophoresis in isotropic electrolytes [7-10] that requires the particles to be asymmetric (such as metal-dielectric Janus spheres), the advantage of the DI-LCEP is in the ability to move perfectly symmetric homogeneous spheres. As compared to the conventional LCEP that moves particles with dipolar director configuration [12,13,14], the advantage of the DI-LCEP is in the ability to move particles that show a higher symmetry of the director in absence of the field. Moreover, in LCEP, the colloids move only parallel to the background director $\hat{\mathbf{n}}_0$, while in DI-LCEP, the direction of motion can be tuned by the electric field.

Steering of colloidal transport is attracting a considerable interest lately. The LCEP mechanism has been demonstrated to control the direction of colloids by patterned surface director in the plane of the cell [11, 21, 22] or even in three-dimensional space, by combining LCEP with linear electrophoresis [23]. Hernàndez-Navarro et al. [13] reported on reconfigurable swarms of asymmetric pear-shaped colloids driven by LCEP and steered by photoactivated photo-switchable surface anchoring. Sahu, Ramaswamy, and Dhara [24] reported on an in-plane omnidirectional transport of metal-dielectric Janus spheres that is based on the asymmetries of both the particles and the surrounding director field; the direction of propulsion is controlled by varying the field frequency and amplitude [24]. In the described DI-LCEP effect, the particle is also steered in the plane of the cell by changing the frequency and voltage of the AC electric field, but the difference is that the particle is a symmetric homogeneous sphere.

The interdependency of the surface properties of the colloids, symmetry of the directron dresses, field parameters such as amplitude and frequency, material parameters of both the liquid crystal electrolyte and the colloids and the direction and speed of the particles driven by DI-LCEP suggests that the described mechanism can bring about many different dynamic scenarios worthy of further studies.

## ACKNOWLEDGMENTS

The work was supported by NSF grant DMR-1905053.